\pdfoutput=1
\documentclass[mathleft,fleqn,%
]{an}
\usepackage{amsmath}
\usepackage{graphicx}
\input macros 
\usepackage{natbib}
\bibpunct{(}{)}{;}{a}{}{,}
\begin{document}

\Pagespan{1}{}
\Yearpublication{2015}%
\Yearsubmission{2015}%
\Month{0}%
\Volume{999}%
\Issue{0}%
\DOI{asna.201400000}%

\title{Chemodynamical modelling of the Milky Way}

\author{James Binney\inst{1}\fnmsep\thanks{Corresponding author:
        {binney@thphys.ox.ac.uk}}
\and  Jason L. Sanders\inst{1,2}
}
\titlerunning{Modelling the Milky Way}
\authorrunning{James Binney \& Jason L. Sanders}
\institute{
Rudolf Peierls Centre for Theoretical Physics, Keble Road, Oxford OX1 3NP
\and 
Institute of Astronomy, Madingley Road, Cambridge CB3 0HA
}

\received{9 October 2015}
\accepted{26 November, 2015}
\publonline{XXXX}


\abstract{Chemodynamical models of our Galaxy that have analytic Extended
Distribution Functions (EDFs) are likely to play a key role in extracting science
from surveys in the era of Gaia.
}

\maketitle

\section{Introduction}
Enormous observational resources are currently being expended on surveys of
our Galaxy, and these expenditures will continue for at least the next five
years. Consequently a major task of contemporary astronomy is to forge the
resulting data into a coherent working model of our Galaxy. The
community hopes to unravel the Galaxy's assembly history, but
such hopes will remain illusory until we have firmly established what's out
there: how is the Galaxy currently structured, and how is it working?

Given the extent to which a star's chemistry must encode the time and place
of its formation, and the way stellar kinematics
encodes the Galaxy's gravitational field and through that the Galaxy's
distribution of matter, chemodynamical models will be central to extracting
science from surveys.
The central role of chemodynamical models is further underlined by three
issues: 

\begin{itemize}

\item The contents of survey catalogues are strongly influenced by
selection effects: catalogues are dominated by what's near or luminous. It's
easier to account for  selection effects by imposing them on mock
observations of a model than to ``correct'' the observational data as was
often done in the past.

\item The majority of stars in any catalogue are close to the survey's
limiting magnitude or resolution. Consequently, observational errors are
small for only a minority of surveyed stars. It is much  easier to impose the
observational errors on mock observations than to correct the data. The
revolution in data gathering over the last couple of decades ensures that
statistical errors are no longer the limiting source of uncertainty in
scientific understanding; we are rather limited
by incomplete understanding of observational errors (``systematic'' errors).

\item A good chemodynamical model summarises the knowledge we have gained
from several different surveys, and indeed tests this knowledge by checking
that conclusions we have drawn from different bodies of data are mutually
consistent. In fact, our ultimate, essentially complete understanding of our
Galaxy will be embodied in the final chemodynamical model, which we will come
to identify with the Galaxy in the same way that in the 19th century our
planet was identified with a school-room globe.

\end{itemize}

\section{What models to use?}

\subsection{Cosmological models}
 Simulations of dark-matter clustering form the backbone of cosmology. The
physics that goes into them is extremely simple, yet it took 15 years for the
simulations to get the basic picture right. For the last 15 years the
community has struggled with simulations that include gas and star formation.
It is now clear that the majority of baryons reside in the intergalactic
medium and without powerful feedback from star formation galaxies become too
luminous and too spheroidal. The effectiveness of feedback must be connected
to the near fractal structure of the ISM. Cosmological simulations won't be
able to resolve this structure in the foreseeable future, so the community is
hunting for an ``effective field theory'' that describes that
$\ga10\pc$-scale impact of this unresolved structure. There is no guarantee
that such an effective theory exists, and certainly we can't use its results
now.  Consequently, we have no ability to compute the consequences of the
precise initial conditions specified by the concordance cosmology. In short
cosmological simulations lack predictive power.

It's hard even to characterise a cosmological simulation: if the equations of
motion are integrated for a few timesteps, the billions of numbers that
specify it change while the
model remains the same. It's a totally non-trivial task to ask if two
simulations differ materially, and if they do, in what way.

Cosmological simulations are computationally costly -- one model typically
requires years of CPU time. Consequently, there's no
realistic prospect of fitting them to observational data. From this it
follows that the insights they provide, though possibly useful, are of an
anecdotal nature. 

\subsection{Oxford models}
 If we are to fit models to data, it's vital to minimise the number of
parameters  required to specify a model.
Over 90\% of the Galaxy's matter is contained in unobserved dark matter and
we have to constrain the distribution of this material through the effect its
gravitational field has on stars and
gas. This we can do only to the extent that the Galaxy is statistically
stationary because {\it any\/} phase-space coordinates of stars are consistent
with {\it any\/} gravitational potential $\Phi$ until you require that after a
dynamical time has elapsed the system has not expanded or collapsed or
otherwise changed its shape to a significant extent. Consequently, we focus
on equilibrium Galaxy models. The Galaxy is not in exact equilibrium, but we
hope to model non-equilibrium phenomena such as spiral structure and the warp by
perturbing an equilibrium model.

By Jeans' theorem the distribution function (\df ) of an equilibrium model can
be taken to be a function $f(I_1,\ldots)$ of the constants of stellar motion.
Most orbits in typical axisymmetric potentials are quasiperiodic in the sense
that Fourier decomposition of the time series $x(t)$ obtained by numerical
integration of the equations of motion contains power only at discrete
frequencies $\omega$ that can be expressed as integer linear combinations
$\omega=n_1\Omega_1+n_2\Omega_2+n_3\Omega_3$ of three fundamental frequencies
$\Omega_i$. This fact guarantees the existence of angle-action coordinates
$(\vtheta,\vJ)$, which are such that the actions $J_i$ are constants of
motion and $\theta_i$ increases linear with time:
$\theta_i(t)=\theta_i(0)+\Omega_it$

The action $J_r$ quantifies excursions in $r$, the action $J_\phi$ is simply
the component $L_z$ of angular momentum about the Galaxy's symmetry axis,
and the action $J_z$ quantifies motion perpendicular to the Galactic plane.
In Oxford we have developed techniques for computing transformations between
action-angle and ordinary $(\vx,\vv)$ coordinates
\citep{BinneyFudge,SandersGF,SandersFudge}. Our techniques
are by no means optimal and we continue to refine them. But they work and
allow us to fit data to models defined by \df s $f(\vJ)$.

\section{Distribution functions}

We think of galaxies as sums of ``components'' or ``populations''. Any
non-negative function $f(\vJ)$ defines an equilibrium dynamical model of a
component. A huge advantage of making the \df\  depend on the actions rather
than energy $E$ is that then adding components is straightforward: you simply
add the corresponding functions of $\vJ$
\[
f(\vJ)=\sum_i f_i(\vJ).
\]
 By contrast, if $E$ is permitted to be an argument of $f$, the system (if
any) generated by the summed \df\  is in no useful sense the sum of its
components because the presence of one component affects the energy scale of
all components, just as it does the energy of a circular orbit of radius $R_0$.
Moreover, finding the potential $\Phi$ that self-consistently corresponds to
a \df\  $f(\vJ)$ is easy \citep{BinneyHenon} but hard when $E$ is an argument of
$f$ \citep[e.g][]{PrendergastTomer}.

\subsection{DF for thin and thick discs}
We use the \df\ 
\[\label{eq:QIDF}
f_\sigma(\vJ)={\Sigma\Omega\nu\over2\pi^2\sigma_r^2\sigma_z^2\kappa}
[1+\tanh(J_\phi/L_0)]\e^{-\kappa J_r^2/\sigma_r^2-\nu J_z/\sigma_z^2}
\]
 of a ``quasi-isothermal'' disc as the building block from which to assemble
realistic models of the Galactic disc(s). In this \df\  $\Omega(J_\phi)$ is the
angular frequency of the circular orbit with angular momentum $J_\phi$, while
$\kappa(J_\phi)$ and $\nu(J_\phi)$ are the radial and vertical epicycle
frequencies of this orbit.  $\Sigma$ also depends on $J_\phi$ and is to
a good approximation the surface density at the radius $\Rc(J_\phi)$ of the
circular orbit. The choice $\Sigma=\Sigma_0\e^{-\Rc/R_\rd}$ gives rise to a
nearly exponential disc. The functions $\sigma_r(J_\phi)$ and
$\sigma_z(J_\phi)$ control the radial and vertical velocity dispersions
within the disc. A disc with roughly radius-independent scale height is
obtained by taking $\sigma_0\propto\e^{-\Rc/R_\sigma}$ with
$R_\sigma\simeq2R_\rd$. The purpose of the  tanh function in eqn
(\ref{eq:QIDF}) is to eliminate counter-rotating orbits. For this purpose,
$L_0$ should not be bigger than the typical angular momentum of a star in the
bulge/bar. We typically adopt $L_0=10\kpc\kms$.

To date we have assumed that the thick disc can be adequately described by a
single quasi-isothermal. This assumption is surely too crude, but so are the
currently available data for the thick disc.

We take the \df\  of the thin disc to be a superposition of quasi-isothermals,
one for the stars of each age. Specifically
\[
f_{\rm thn}(J_r,J_z,L_z)
=\int_0^{\tau_m}\rd\tau\,\Gamma(\tau)f_{\sigma_\tau}(\vJ),
\]
 where
\[
\Gamma(\tau)={\e^{\tau/t_0}\over t_0(\e^{\tau_m/t_0}-1)}
\]
 encodes the history of star formation 
 and  the velocity-dispersion functions increase with age:
\[\begin{split}
\sigma_r(J_\phi,\tau)&=\sigma_{r0}\left({\tau+\tau_1\over\tau_{\rm
m}+\tau_1}\right)^{\beta_r}\e^{(R_0-\Rc)/R_\sigma}\cr
\sigma_z(J_\phi,\tau)&=\sigma_{z0}\left({\tau+\tau_1\over\tau_{\rm
m}+\tau_1}\right)^{\beta_z}\e^{(R_0-\Rc)/R_\sigma}.
\end{split}
\]
 We typically adopt $\beta_r=0.33$ and $\beta_z=0.4$ so the velocity
dispersions of a cohort increase rapidly at first and slower later on.
$\tau_{\rm m}\simeq10\Gyr$ is the age of the thin disc, and $\tau_1$ is
chosen such that the velocity dispersion of stars at birth is $\sim7\kms$.
With these choices one obtains a disc in which the youngest stars are
confined to a very thin disc in which the radial and vertical velocity
dispersions are small, and the oldest stars form a thicker disc in which
stars move on more eccentric orbits. 

\subsection{DF for stellar and dark halo}

\citet{Posti2015} showed that self-consistent systems very similar to the
Hernquist, NFW and Jaffe spheres can be generated by particular instances of
the \df\ 
\[\label{eq:Posti}
f(\vJ)={M_0\over J_0^3}
{[1+J_0/h(\vJ)]^{(6-\alpha)/(4-\alpha)]}\over[1+g(\vJ)/J_0]^{2\beta-3}}.
\]
 Here $h(\vJ)$ and $g(\vJ)$ are homogeneous functions of degree one, i.e.,
$h(\gamma\vJ)=\gamma h(\vJ)$, etc. For $|\vJ|\ll J_0$ $h$ dominates $f$ so
$h$ controls the central structure of the resulting model. For $|\vJ|\gg
J_0$, $g$ dominates $f$, so $g$ controls the structure of the model's halo.
By varying the way $h$ and $g$ depend on the individual components of $\vJ$,
the model can be made prolate or oblate and its velocity distribution can be
made tangentially or radially biased. We take the \df s of the stellar and
dark halos to be instances of the \df\ (\ref{eq:Posti}).

\section{Metallicity-blind models}

\citet{Binney2010,Binney2012} examined models defined by a disc \df\  and a specified
gravitational potential that was a sum of the contributions from a
double-exponential stellar disc, a gas disc, and simple spheroidal models of
the bulge and the dark halo. A success of this
work was that it indicated that the accepted value of the azimuthal component $V$ of
the solar velocity with respect to the Local Standard of Rest was many sigma
too small. \citet{SchoenrichBD} subsequently
explained how the metallicity gradient in the disc led \citet{DehnenB} to
extract an erroneous value of $V$ from the Hipparchos data.

\citet{Binney2012} had fitted the disc \df\  to the Geneva-Copenhagen survey
\citep[GCS:][]{Nordstrom2004,Holmberg2009}, which contains only nearby stars ($s\la120\pc)$.
\citet{Binney2014} compared the predictions of this \df\  for the contents of the
RAVE survey \citep{Steinmetz2006,Kordopatis2013}, which extends to
$s\ga2\kpc$. The predictions for $v_z$ were extremely successful. Those for
$v_R$ and $v_\phi$ quite successful, but far from the plane the \df\  predicted
values of $\sigma_R$ that were too small. Given the small contribution that
the thick disc makes to the GCS, these shortcomings in the predictions for
RAVE were to be expected. \citet{Piffl2014} fitted the \df s of the disc and
stellar halo to the RAVE data and showed that excellent fits could be
obtained.

In the work just described the potential was generated by analytic models of
the Galaxy's components rather than by the mass distributions predicted by
the \df. \citet{PifflPB} computed the observable properties of the model
produced by the \df\ of \citet{Piffl2014} when the dark matter was also
predicted by a \df\ rather than a specified density distribution. The chosen
halo \df\ was the one that in isolation generates the density distribution
that \citet{Piffl2014} adopted for the dark halo. When this same \df\ is used
to evaluate the density of dark matter in the presence of the disc, the dark
halo is more centrally concentrated than \citet{Piffl2014} assumed.
Consequently, the self-consistent model that \citet{PifflPB} recovered had a
circular-speed curve that exceeded observational limits at $R\la4\kpc$. 

\citet{BinneyP} investigated whether  this failure could be avoided by
modifying the \df s of the disc and dark halo. They found that a model could
be found that satisfied all the observational constraints considered by
\citet{Piffl2014} but the model generates too little optical depth to
microlensing of bulge stars. The core problem is that an adiabatically
compressed dark halo is too centrally concentrated. If such a dark halo is to
contribute to the circular speed at $R_0$ to the required extent, it will
contribute excessively to the rotation curve at $R\sim3\kpc$. This is the
first clear observational indication that the Galaxy's baryons have not accumulated
adiabatically.

\section{Extended DFs}

In all the above work the \df\  was fitted
to catalogues without regard for the
selection function. That is, the catalogue was taken to define a
complete population that could be consistently described by a \df. Actually,
the population picked up by a survey at small distances differs from that
selected at large distances because the nearby stars includes stars of low
luminosity that will not be picked up at greater distances.

To model surveys with significant depth consistently, the model has to take
cognisance of the factors that determine stellar luminosity and colour: mass,
age, and metallicity. The \df s introduced by \citet{Binney2010} recognised
mass to the extent that a universal IMF was implicitly assumed, and age
featured explicitly. But in comparing the resulting predictions with data
this information was only used in a very simplistic way.  \citet[][hereafter
SB15b]{SandersEDF}
took a significant step forward by extending the \df\  to include metallicity
[Fe/H] and in comparisons with data used isochrones to exploit properly the
information inherent in mass, age and metallicity. When metallicity features
in the \df , we say that the model has an \edf\  for Extended \df .

SB15b derived a form for the Galaxy's \edf\  by constructing
analytic approximations to distributions that arise in the detailed
chemodynamical models of \citet{SchoenrichB}. A time $\tau$ in the past the
iron abundance $F\equiv\hbox{[Fe/H]}$ of the ISM at radius $R$ is taken to be
\[
F(R,\tau)=F_{\rm m}+\left[F(R)-F_{\rm m}\tanh\left({\tau_{\rm m}-\tau\over\tau_{\rm
F}}\right)\right],
\]
 where $\tau_{\rm m}\simeq12\Gyr$, $\tau_{\rm F}\simeq 3\Gyr$ and
\[\label{eq:EDF2}
F(R)=\left(1-\exp\left[{-F_R(R-R_{\rm F})\over F_{\rm m}}\right]\right)F_{\rm m}.
\]
 According to these formulae, the abundance rises at a given radius from a
pre-enrichment level $F_{\rm m}$ to its current value $F(R)$ on a short
timescale $\tau_{\rm F}$. $R_{\rm F}$ is the radius at which
[Fe/H] now vanishes, so $R_{\rm F}\simeq R_0$. For $R\simeq R_{\rm
F}$ the argument of the exponential in eqn (\ref{eq:EDF2}) is small, and
expanding the exponential we see that $F_{\rm R}$ is the metallicity gradient
at $R_{\rm F}$. Consequently, it is negative. It follows that $F_{\rm m}$ is
the current iron abundance far from the Galactic centre. Once $R_{\rm F}$,
$F_{\rm m}$ and $F_{\rm F}$ have been chosen, the iron abundance at the
Galactic centre is
determined.

SB15b took the age distribution of thin-disc stars to be
\[
\Gamma(\tau)={1\over G_0}\exp\left({\tau\over\tau_{\rm f}}-{\tau_{\rm
s}\over\tau_{\rm m}-\tau}\right).
\]
 When $\tau$ is only slightly smaller than the age $\tau_{\rm m}$ of the
Galaxy, the second term in the exponential is large, so the star-formation
rate is small. So at early times the SFR rises rapidly. It peaks and then
declines as the first term in the exponential takes over. Our work with plain
\df s relied on the first term in the exponential, thus an SFR that always
decreases. As soon as we tried to fit the local metallicity distribution, we found
it essential to add the second term, which encodes an early rise in the SFR
to a peak about $2\Gyr$ after the Galaxy started to form.

Given the metallicity of stars born at each time and radius, the density
$f(\vJ,\tau,F)$ of
stars with given actions age and metallicity would follow if one had the
Green's function of the equation
\[
{\p f\over\p t}={\p\over\p\vJ}\cdot\left(-\vD^{(1)}f+\vD^{(2)}\cdot{\p
f\over\p\vJ}\right)
\]
 that governs the diffusion of stars through action space. Here $\vD^{(1)}$
and $\vD^{(2)}$ are the first- and second-order diffusion coefficients. A
suitable Ansatz for the
Green's function for diffusion in $J_\phi$ is
\[
G(J_\phi,J_\phi',t)=\sqrt{{\tau_{\rm m}/t\over2\pi\sigma^2_{L0}}}
\exp\left[-{(J_\phi-J_\phi'-D_\phi^{(1)}t)^2\over2\sigma^2_{L0}t/\tau_{\rm
m}}\right],
\]
 where $\sigma_{L0}$ is a constant related to the magnitude of
$D_{\phi\phi}^{(2)}$. Since the angular momentum of the entire disc
does not change as stars diffuse in angular momentum, we require
$\int\rd^3\vJ\,J_\phi (\p f/\p t)=0$, which implies that
\[
D_\phi^{(1)}=-{\sigma_{L0}^2\over2\tau_{\rm m}V_cR_\rd}.
\]

\begin{figure*}
\centerline{\includegraphics[height=0.23\textheight, bb = 8 8 216
173]{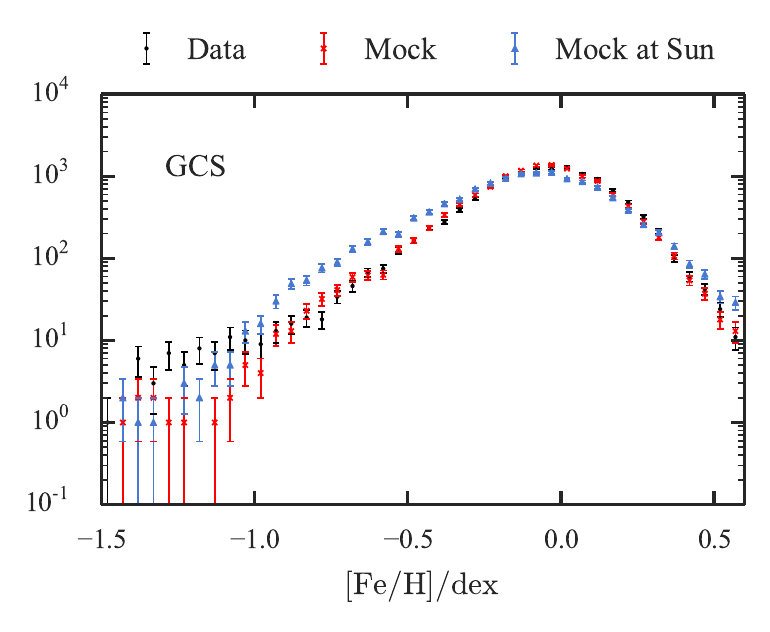}\quad
\includegraphics[height=0.23\textheight, bb = 8 8 221
176]{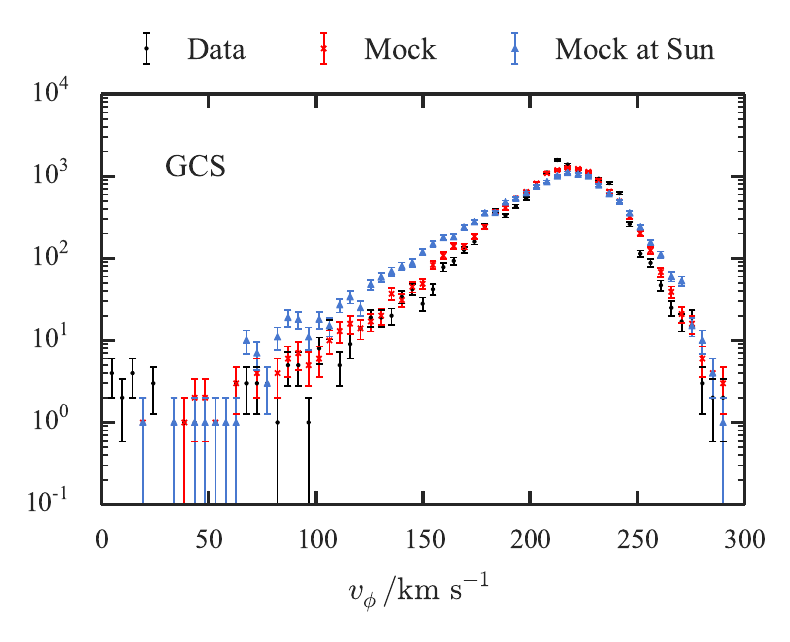}}
\caption{The fit to GCS data provided by the \edf. Black points show the data,
and red points the fit obtained. The pale blue points show the predictions of
the \edf\  when no account is taken of the selection bias of the
GCS.}\label{fig:GCSfit}
\end{figure*}
\begin{figure*}
\centerline{\includegraphics[height=0.23\textheight, bb = 8 8 216
173]{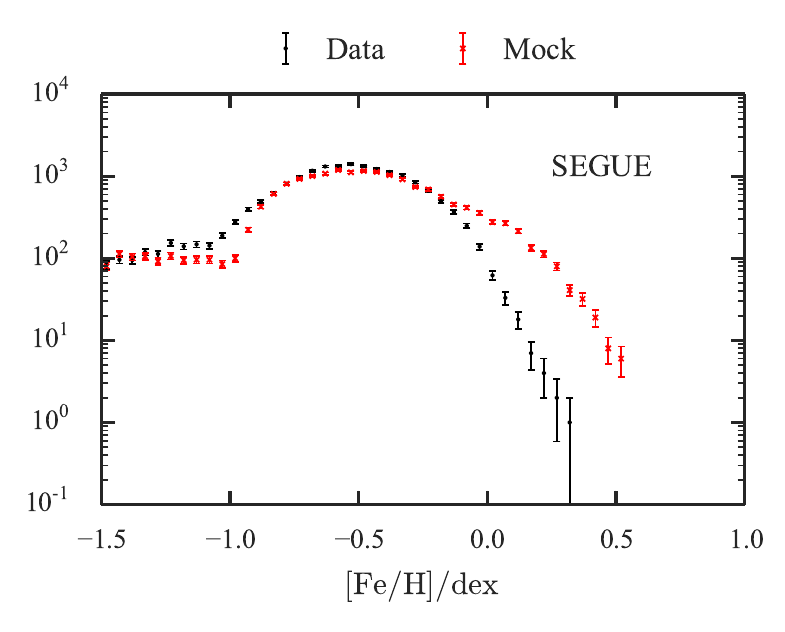}\quad
\includegraphics[height=0.23\textheight, bb = 8 8 221
176]{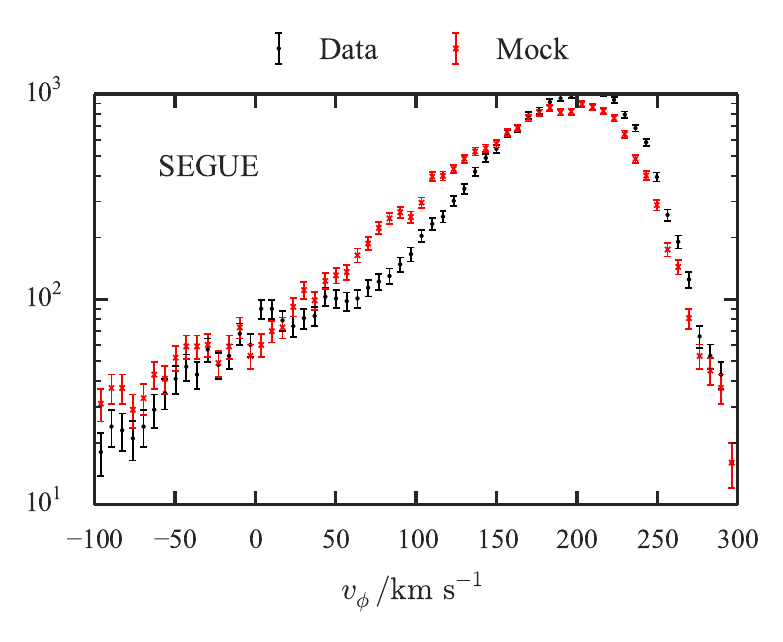}}
\caption{Predictions of the \edf\  for the statistics of G dwarfs from SEGUE.
Black points show the data,
and red points the predictions.}\label{fig:SEGUE} 
\end{figure*}

The red points in Fig.~\ref{fig:GCSfit} show two of the excellent fits to the
GCS catalogue (black points) that we obtain using this \edf. The only
statistically significant defect is a slight under-prediction of the number
of stars with $v_z>50\kms$. This defect might reflect a poor choice of Galaxy
potential. Since the parameters of the \df\  were adjusted to ensure that the
stellar density profile $\rho(z)$ above the Sun agreed with the data points
of \citet{GilmoreR1983}, if our potential under-estimates the vertical force
$K_z$, the distribution in $v_z$ that it provides will be too narrow.

The fit calls for the thin disc to have a significantly longer scalelength
$R_\rd=3.45\kpc$ than that of the thick disc, $R_\rd=2.31\kpc$. The local
metallicity gradient is $F_{\rm R}=-0.064\,\hbox{dex}\kpc^{-1}$ and the timescale of metallicity
increase is $\tau_{\rm F}=3.2\Gyr$. The early star-formation
timescale is $\tau_s=0.43\Gyr$, and the timescale of subsequent decay is
$\tau_{\rm f}=8\Gyr$. 

The pale blue points in Fig.~\ref{fig:GCSfit} show the predictions of the \edf\ 
when the selection function is set to unity. Then the \edf\  predicts increased
numbers of stars with low [Fe/H] on eccentric orbits (which imply low
$v_\phi$ and high $v_R$). The actual selection function is biased towards
young stars, which tend to have higher metallicities and more circular
orbits. In earlier work we, in effect, adjusted the \df\  to make the blue
points coincide with the data. Consequently, we then derived values of the
velocity-dispersion parameters $\sigma_{r0}$ and $\sigma_{z0}$ that are too
small.  Specifically, fitting with the proper selection function yields
$(\sigma_{r0},\sigma_{z0})=(48.3,30.7)\kms$ for the thin disc, and
$(50.5,51.3)\kms$ for the thick disc, whereas using the same potential but
neglecting the selection function, \citet{Binney2012} obtained $(42.2,20.3)$
for the thin disc 
and $(26.3,34.0)\kms$ for the thick disc.

It's interesting to compare the {\it predictions} of the \edf\ fitted to the
GCS with data for much more distant G-dwarfs from the SEGUE survey
\citep{Ahn2014}. The red points in Fig.~\ref{fig:SEGUE} show these
predictions. The main problem is that the model predicts more stars with
$\hbox{[Fe/H]}>0$ than the data show. We suspect the problem lies
with the observed metallicities rather than the model.

\section{Conclusion}

Models with analytic \df s $f(\vJ,\hbox{Fe/H})$ can provide remarkably good
fits to current data and we argue are the key to extracting a coherent
picture of our Galaxy from survey data. Including [Fe/H] in the \df\  enables
one to handle selection functions, and doing so is very important. The EDF we
use is inspired by an evolutionary scenario but it is a valid description of
the present-day Galaxy regardless of the truth of the scenario. Conversely,
the success of the EDF does not establish the validity of the scenario.

\bibliographystyle{mn2e}
\bibliography{refs}

\end{document}